\documentclass[twocolumn,preprintnumbers,aps,showpacs,superscriptaddress, pra]{revtex4-1}
\usepackage{palatino}
\usepackage{natbib}
\usepackage[latin1]{inputenc}
\usepackage{epsf}
\usepackage{amsmath,amssymb}
\usepackage{latexsym}
\usepackage{calc}
\usepackage{color}
\usepackage{epsfig}

\usepackage{hyperref}

\newcommand{\ben}{\begin{equation}}
\newcommand{\een}{\end{equation}}
\newcommand{\bea}{\begin{eqnarray}}
\newcommand{\eea}{\end{eqnarray}}

\def\bA{{\bf A}}

\def\dulR{{\underline{\underline{\bf R}}}}
\def\dulr{{\underline{\underline{\bf r}}}}

\begin{document}
\title{Bohmian mechanics in the exact factorization of electron-nuclear wavefunction} 
\author{Yasumitsu Suzuki}
\affiliation{Department of Physics, Tokyo University of Science, 1-3 Kagurazaka, Shinjuku-ku, Tokyo 162-8601, Japan}  
\author{Kazuyuki Watanabe}
\affiliation{Department of Physics, Tokyo University of Science, 1-3 Kagurazaka, Shinjuku-ku, Tokyo 162-8601, Japan}  
\date{\today}
  
\begin{abstract}
The exact factorization of an electron-nuclear wavefunction [A. Abedi, N. T. Maitra, and 
E. K. U. Gross, Phys. Rev. Lett. 105, 123002 (2010)]
allows us to define the rigorous nuclear time-dependent Schr\"{o}dinger equation (TDSE) with
a time-dependent potential-energy surface (TDPES) that fully accounts for the coupling 
to the electronic motion and drives the nuclear wavepacket dynamics. Here, we study 
whether the propagation of multiple classical trajectories can reproduce the quantum 
nuclear motion in strong-field processes when their motions are governed by the quantum 
Hamilton-Jacobi equation derived by applying Bohmian mechanics to this exact nuclear TDSE. 
We demonstrate that multiple classical trajectories propagated by the force from the gradient 
of the exact TDPES plus the Bohmian quantum potential can reproduce the strong-field 
dissociation dynamics of a one-dimensional model of the H$_2^+$ molecule. Our results show 
that the force from the Bohmian quantum potential plays a non-negligible role in yielding 
quantum nuclear dynamics in the strong-field process studied here, where ionization 
and/or splitting of nuclear probability density occurs.
\end{abstract}

\maketitle

\section{Introduction}
The coupling of electronic and nuclear motions plays a significant role in many fascinating 
scientific phenomena, e.g., photoreactions
\cite{photo1,photo2,photo3}, molecular electronics
\cite{molel1,molel2,molel3}, 
and strong-field processes
\cite{SFEN1,SFEN2,SFEN3,SFEN4,SFEN5}.
In particular, the study of strong-field processes has been one of the most dynamic research 
areas in the past few decades with the advent of femtosecond and attosecond technology
\cite{atto0,atto1,atto2,atto3,atto4,atto5}.
Irradiation of atoms and molecules by intense laser pulses gives rise to highly nonlinear effects
\cite{SFP1,SFP2,SFP3,SFP4,SFP5,SFP6,SFP7,SFP8,SFP9} such as above-threshold ionization or 
dissociation, Coulomb explosion, or high-harmonic generation. To understand the mechanisms 
of any nonlinear molecular phenomenon and non-adiabatic reactions, and to move the technology 
further forward, it is essential to be able to correctly describe any type of coupled 
electron-nuclear dynamics. Developing such a theoretical tool has been one of the 
biggest issues in theoretical physics and chemistry. Numerous studies have been conducted 
and many sophisticated methods have been developed
\cite{enpaper0, enpaper01, enpaper1, enpaper2, enpaper3, enpaper4,
enpaper5, enpaper6, enpaper7,enpaper8,enpaper9,enpaper10,enpaper11}, 
among them, multiple-spawning method
\cite{spawn1,spawn2},
multiconfiguraton time-dependent Hartree method
\cite{MCTDH1,MCTDH2,MCTDH3,MCTDH4}, and
nonadiabatic Bohmian dynamics method
\cite{NBD1,NBD2,Conditional} are the methods that retain
a quantum description of the nuclei and simulate nonadiabatic electron-nuclear dynamics
very accurately.
However, they incur huge computational cost when applied to systems with many atoms. 
Moreover, inclusion of a large number of electronic states is required when higher-intensity 
fields exist, and ionization processes are very difficult to treat within these approaches.
Therefore, alternative approaches have also been developed extensively, which have significantly 
reduced the computational cost. One of the most widely used approaches is mixed 
quantum-classical (MQC) approximation, where the nuclei are treated as classical particles, 
while the electrons are treated quantum mechanically. Among these, the Ehrenfest
\cite{Ehrenfest1, Ehrenfest2, Ehrenfest3} 
and trajectory surface hopping (TSH)
\cite{TSH1,TSH2,TSH3,TSH4,TSH5,TSH6,TSH7,TSH8} 
methods are the most widely used, and have been employed in many studies. However, both 
Ehrenfest and TSH have certain discrepancies that arise from the fact that, in both methods, 
the forces acting on classical nuclei and the potential acting on electrons are derived with 
approximations. There are ongoing intensive efforts to improve these approaches
\cite{dec1,dec2,dec3,dec4,dec5,dec6,dec7}.

Recently, a new approach to the coupled electron-nuclear motion, the so-called exact 
factorization of the electron-nuclear wavefunction
\cite{AMG,AMG2,*AMG2C,*AMG2R,SAMYG}, has been proposed. This method provides a new 
route to go beyond the Born-Oppenheimer 
(BO) approximation
\cite{bBO1,bBO2,bBO3,bBO4}
and to study the force acting on the classical nuclei
\cite{AAYG,AAYG2,AAYMMG},
and then to develop a rigorous MQC method
\cite{MQC1,MQC2,MQC3,CT1,CT2}.
In this framework, the full wavefunction is written as the product of a nuclear wavefunction 
and conditional electronic wavefunctions, which parametrically depend on the nuclear 
configuration. The coupled equations drive the dynamics of these two components, and 
the motion of the nuclear wavefunction is governed by a single time-dependent Schr\"{o}dinger 
equation (TDSE), which contains a time-dependent potential-energy surface (TDPES) and a 
time-dependent vector potential. Since this nuclear wavefunction provides the exact nuclear 
and current densities, the TDSE that it satisfies has been identified as the exact nuclear 
TDSE. The presence of a single exact nuclear TDSE has been found to be very useful in 
developing the MQC approach systematically
\cite{AAYG,AAYG2,AAYMMG,MQC1,MQC2,MQC3,CT1,CT2}.

In previous studies, the features of the TDPES in a one-dimensional nonadiabatic 
electron-transfer model system have been fully analyzed
\cite{AAYG,AAYG2,Interference}.
Indeed, it has been shown that evolving an ensemble of classical nuclear trajectories 
using the force determined from the gradient of the TDPES reproduces the nuclear wavepacket 
dynamics very well
\cite{AAYMMG}. 
These led to the idea of developing the MQC method based on the TDPES and multiple 
trajectories. Recently, a novel MQC algorithm - the coupled-trajectory (CT) MQC algorithm
\cite{CT1, CT2} - has been proposed and shown to be able to accurately simulate the 
coupled electron-nuclear dynamics in a one-dimensional field-free process.

On the other hand, the features of the TDPES under external fields have also been studied
\cite{AMG,AMG2,*AMG2C,*AMG2R,SAMYG,localization,EAM}.
We proposed the {\it reverse} factorization
\cite{SAMYG}, which allows us to define the exact electronic TDSE 
and the exact electronic TDPES. 
These are found to be very useful for exploring the mechanism of electron dynamics under an 
external field
\cite{SAMYG, EAM}.
Furthermore, we have recently studied the nuclear TDPES in laser-induced electron 
localization in the H$_2^+$ molecule
\cite{localization}, and showed that the propagation of an ensemble of classical trajectories using the 
gradient of the TDPES yields nuclear density dynamics that are very similar to the exact 
quantum nuclear ones. This result encourages the idea of developing the MQC dynamics method 
for the strong-field processes as well. This would be useful since none of the methods that 
presently exist can accurately simulate the coupled electron-nuclear dynamics of medium- and 
large-sized systems under a strong field.

However, it is still not clear whether the gradient of the TDPES can reproduce the quantum 
nuclear dynamics in strong-field processes such as strong-field ionization and dissociation, 
where the quantum effects of the nuclei are significant. In fact, the gradient of the TDPES 
is not exactly the same as the force that appears in the quantum Hamilton-Jacobi equation 
derived by applying the Bohmian mechanics approach to the exact nuclear TDSE
\cite{AAYMMG};
it lacks the force from the so-called Bohmian quantum potential. In previous studies
\cite{AMG,AMG2,*AMG2C,*AMG2R}, 
it was shown that a single classical trajectory evolved by the gradient of the TDPES does 
not yield the correct time evolution of the mean nuclear distance in strong-field dissociation 
of the one-dimensional H$_2^+$ molecular model. The question then arises as to whether multiple 
classical trajectories evolved by the quantum Hamilton-Jacobi equation derived from the exact 
nuclear TDSE give the correct quantum nuclear dynamics in strong-field processes.

In this paper, we show that multiple classical trajectories propagated by the gradient of the 
TDPES {\it plus} Bohmian quantum potential can reproduce quantum nuclear dynamics in the strong-field 
processes: it produces the correct dissociation dynamics and splitting of nuclear probability 
density in the one-dimensional H$_2^+$ molecular model. The Bohmian quantum potential is found to play 
a non-negligible role in giving the correct nuclear dynamics for the present strong-field processes, 
where ionization and/or splitting of the nuclear wavepacket occur/s. Our findings provide a useful 
basis toward the development of the MQC method for strong-field processes.

The rest of the paper is organized as follows. In section II, we briefly review the concepts of the 
exact factorization of the full electron-nuclear wavefunction and the quantum Hamilton-Jacobi equation 
derived from the exact nuclear TDSE. There, we show the exact force acting on the classical nuclei and 
its relationship with the gradient of the TDPES. In section III, we first describe our model system of 
strong-field dissociation of H$_2^+$, and then show the quantum potential in this system together with 
the exact TDPES. We then propagate multiple classical trajectories with the force from the 
gradient of the TDPES plus quantum potential and demonstrate that it perfectly reproduces the 
quantum nuclear dynamics in strong-field processes. We also show the role of the quantum 
potential by showing the dynamics propagated only by the gradient of the TDPES. In section IV, 
we summarize the results and speculate on future directions.

\section{THEORY}

In Ref.
\cite{AMG,AMG2,*AMG2C,*AMG2R}, it was shown that the 
full electron-nuclear wavefunction 
$\Psi(\dulR,\dulr, t)$
that solves the TDSE
$
\hat{H}\Psi(\dulR,\dulr,t)=i\partial_t\Psi(\dulR,\dulr, t)
$
can be factorized exactly to the single product
\ben
\Psi(\dulR,\dulr, t)=\chi(\dulR,t)\Phi_\dulR(\dulr,t)
\label{eqn: factorization}
\een
of the nuclear wavefunction 
$\chi(\dulR,t)$
 and the
electronic wavefunction 
$\Phi_\dulR(\dulr,t)$
that 
parametrically depends on the
nuclear positions 
$\dulR$
and satisfies the partial normalization condition 
\ben
\int d\dulr |\Phi_\dulR(\dulr,t)|^2=1
\een
for any 
$\dulR$ and $t$.
Throughout this paper, 
$\dulR$ and $\dulr$ 
collectively represent the sets of 
nuclear and electronic coordinates, respectively
(i.e., $\dulR \equiv \{ {\bf R}_1,{\bf R}_2,\cdots,{\bf R}_{N_n} \}$
and $\dulr \equiv \{ {\bf r}_1,{\bf r}_2,\cdots,{\bf r}_{N_e} \}$)
, and 
atomic units are used unless stated otherwise.
The complete molecular Hamiltonian is
\ben
\hat{H} = \hat{T}_{\rm n}(\dulR)+ \hat{V}^{\rm n}_{\rm ext}(\dulR,t) +\hat{H}_{\rm BO}(\dulr,\dulR)
+\hat{v}^{\rm e}_{\rm ext}(\dulr,t),
\een
and 
$\hat{H}_{\rm BO}(\dulr,\dulR)$
is the  BO electronic Hamiltonian,
\ben
\hat{H}_{\rm BO} = \hat{T}_{\rm e}(\dulr)+ \hat{W}_{\rm ee}(\dulr) 
+\hat{W}_{\rm en}(\dulr,\dulR)+\hat{W}_{\rm nn}(\dulR),
\een 
where 
$\hat{T}_{\rm n}=-\sum_{\alpha=1}^{N_{\rm n}}\frac{\nabla^2_\alpha}{2M_\alpha}$
and 
$\hat{T}_{\rm e}=-\sum_{j=1}^{N_{\rm e}}\frac{\nabla^2_j}{2}$
are the
nuclear and electronic kinetic energy operators,
$\hat{W}_{\rm ee}$,
$\hat{W}_{\rm en}$ and $\hat{W}_{\rm nn}$ 
are the electron-electron, electron-nuclear and
nuclear-nuclear interactions, and
$\hat{V}^{\rm n}_{\rm ext}(\dulR,t)$ and
$\hat{v}^{\rm e}_{\rm ext}(\dulr,t)$ 
are time-dependent (TD) external
potentials acting on the nuclei and electrons, respectively.

The stationary variations of the quantum mechanical action with respect to 
$\Phi_\dulR(\dulr,t)$ and $\chi(\dulR,t)$
under the normalization condition of 
$\Phi_\dulR(\dulr,t)$ lead to the 
following equations of motion for 
$\chi(\dulR,t)$ and $\Phi_\dulR(\dulr,t)$
\cite{AMG,AMG2,*AMG2C,*AMG2R}:
\ben
\begin{split}
\left(\hat{H}_{\rm BO}(\dulr,\dulR)+\hat{v}^{\rm e}_{\rm ext}(\dulr,t)
+\hat U_{\rm en}^{\rm coup}[\Phi_\dulR,\chi]-\epsilon(\dulR,t)\right)
\Phi_{\dulR}(\dulr,t)\\
=i\partial_t \Phi_{\dulR}(\dulr,t) 
\end{split}\label{eqn: exact electronic eqn}
\een
\ben
\begin{split}
\left[\sum_{\alpha=1}^{N_{\rm n}} \frac{\left[-i\nabla_\alpha+\bA_\alpha(\dulR,t)\right]^2}{2M_\alpha} 
+\hat{V}^{\rm n}_{\rm ext}(\dulR,t) +
\epsilon(\dulR,t)\right]\chi(\dulR,t)\\
=i\partial_t \chi(\dulR,t) \label{eqn: exact nuclear eqn}.
\end{split}
\een
Here,
$\epsilon(\dulR,t)$ 
is the exact nuclear TDPES
\begin{equation}\label{eqn: tdpes}
\epsilon(\dulR,t)=\left\langle\Phi_\dulR(t)\right|\hat{H}_{\rm BO}+\hat{v}^{\rm e}_{\rm ext}(\dulr,t)
+\hat U_{\rm en}^{\rm coup}-i\partial_t\left|
\Phi_\dulR(t)\right\rangle_\dulr,
\end{equation}
$\hat U_{\rm en}^{\rm coup}[\Phi_\dulR,\chi]$
is the electron-nuclear coupling operator,
\begin{align}
\hat U_{\rm en}^{\rm coup}&[\Phi_\dulR,\chi]=\sum_{\alpha=1}^{N_{\rm n}}\frac{1}{M_\alpha}\left[
\frac{\left[-i\nabla_\alpha-\bA_\alpha(\dulR,t)\right]^2}{2} \right.\label{eqn: enco} \\
& \left.+\left(\frac{-i\nabla_\alpha\chi}{\chi}+\bA_\alpha(\dulR,t)\right)
\left(-i\nabla_\alpha-\bA_{\alpha}(\dulR,t)\right)\right],\nonumber
\end{align}
and $\bA_{\alpha}\left(\dulR,t\right)$ 
is the TD vector potential,
\begin{equation}\label{eqn: vector potential}
\bA_{\alpha}\left(\dulR,t\right) = \left\langle\Phi_\dulR(t)\right|-i\nabla_\alpha\left.\Phi_\dulR(t)
\right\rangle_\dulr.
\end{equation}
The symbol
$\left\langle\,\,\cdot\,\,\right\rangle_\dulr$
indicates an integration over electronic coordinates only. 
The partial normalization condition of $\Phi_\dulR(\dulr,t)$ makes the 
factorization~(\ref{eqn: factorization}) unique up to within a
$(\dulR,t)$-dependent gauge transformation, 
$
\chi(\dulR,t)\rightarrow\tilde\chi(\dulR,t)=e^{-i\theta(\dulR,t)}\chi(\dulR,t)$ and $ 
\Phi_\dulR(\dulr,t)\rightarrow\tilde\Phi_\dulR(\dulr,t)=e^{i\theta(\dulR,t)}\Phi_\dulR(\dulr,t)$.
Eqs. (\ref{eqn: exact electronic eqn}) and (\ref{eqn: exact nuclear eqn}) are form invariant under 
this transformation while 
the scalar potential and the vector potential transform as  
$
\tilde{\epsilon}(\dulR,t) = \epsilon(\dulR,t)+\partial_t\theta(\dulR,t)$ and
$\tilde{\bf A}_{\alpha}(\dulR,t) = {\bf A}_{\alpha}(\dulR,t)+\nabla_\alpha\theta(\dulR,t)$.

The equation for $\chi(\dulR,t)$, Eq.~(\ref{eqn: exact nuclear eqn}), has the form of a Schr\"odinger equation.
Note that $\chi(\dulR,t)$ can be interpreted 
as the exact nuclear wave-function since it leads to an $N$-body nuclear density, 
$\Gamma(\dulR,t)=\vert\chi(\dulR,t)\vert^2,$ and an $N$-body current density, 
${\bf J}_\alpha(\dulR,t)=\frac{1}{M_\alpha}\Big[\mbox{Im}(\chi^*(\dulR,t)\nabla_\alpha\chi(\dulR,t))+
\Gamma(\dulR,t){\bf A}_\alpha(\dulR,t)\Big],$  which yields the true nuclear 
$N$-body density and current density obtained from the full
wavefunction $\Psi(\dulR,\dulr,t)$
\cite{AMG2,*AMG2C,*AMG2R}. 
Therefore, the equation~(\ref{eqn: exact nuclear eqn}) can be regarded
the {\it exact} nuclear TDSE.

Having a single exact TDSE for the nuclear subsystem, it is possible to consider its hydrodynamic 
reformulation~\cite{AAYMMG} using the approach of Bohmian mechanics
\cite{Bohm0, Bohm1, Bohm2},
to study how the exact force acting on the classical nuclei can be defined. To this end, the polar 
forms of the wavefunction
$\chi(\dulR,t)=\vert\chi(\dulR,t)\vert e^{iS(\dulR,t)}$
($\vert\chi(\dulR,t)\vert$ and $S(\dulR,t)$ are real-valued
amplitude and action functions, respectively) are substituted into the exact nuclear TDSE~(\ref{eqn: exact nuclear eqn}).
Here, to easily find the exact force acting on the classical nuclei, we set the gauge of the 
wavefunction such that the vector potential
 ${\bf A}_\alpha(\dulR,t)$ is always zero.
Note that whenever the vector potential is curl-free ($\nabla_\alpha\times{\bf A}_\alpha(\dulR,t)=0$),
the gauge can be chosen such that ${\bf A}_\alpha(\dulR,t)$ is zero. 
Whether and under which conditions $\nabla_\alpha\times{\bf A}_\alpha(\dulR,t)=0$ is currently
under investigation
\cite{Gphase1, Gphase2}.
Under this choice of the gauge, the exact nuclear TDSE~(\ref{eqn: exact nuclear eqn}) can be written as
\ben
\left[-\sum_{\alpha=1}^{N_n} \frac{\nabla_\alpha^2}{2M_\alpha} +
\epsilon(\dulR,t)\right]\chi(\dulR,t)
=i\partial_t \chi(\dulR,t) \label{eqn: exact nuclear eqn2}.
\een
Here we include $\hat{V}^{\rm n}_{\rm ext}(\dulR,t)$ in the exact TDPES $\epsilon(\dulR,t)$.
Substituting $\chi(\dulR,t)=\vert\chi(\dulR,t)\vert e^{iS(\dulR,t)}$ into Eq.~(\ref{eqn: exact nuclear eqn2}),
the following two coupled equations are obtained
\cite{AAYMMG}:
\ben
\begin{split}
\sum_{\alpha=1}^{N_{\rm n}}\frac{(\nabla_\alpha S(\dulR,t))^2}{2M_\alpha}+\epsilon(\dulR,t)
-\sum_{\alpha=1}^{N_{\rm n}}\frac{1}{2M_\alpha}\frac{\nabla_\alpha^2 \vert\chi(\dulR,t)\vert}{\vert\chi(\dulR,t)\vert}\\
=-\frac{\partial S(\dulR,t)}{\partial t}
\label{eqn: QHJeqn}
\end{split}
\een
and
\ben
\begin{split}
-\sum_{\alpha=1}^{N_{\rm n}}\frac{\nabla_\alpha S(\dulR,t)\cdot\nabla_\alpha \vert\chi(\dulR,t)\vert}{M_\alpha}
-\sum_{\alpha=1}^{N_{\rm n}}\frac{\vert\chi(\dulR,t)\vert\nabla_\alpha^2 S(\dulR,t)}{2M_\alpha}\\
=\frac{\partial \vert\chi(\dulR,t)\vert}{\partial t}.
\label{eqn: continuity}
\end{split}
\een
Equation (\ref{eqn: QHJeqn}) and (\ref{eqn: continuity}) are the hydrodynamic formulation of 
the exact nuclear TDSE~(\ref{eqn: exact nuclear eqn2}).
Equation~(\ref{eqn: QHJeqn}) can be regarded as the {\it exact} quantum Hamilton-Jacobi equation,
while~(\ref{eqn: continuity}) produces the continuity equation.
Identifying $\nabla_\alpha S(\dulR,t)$ as a momentum ${\bf P}_\alpha$ of a classical trajectory,
(\ref{eqn: QHJeqn}) can be solved by propagating an ensemble of classical
trajectories that obey the following Newton's equations:
\ben
\begin{split}
\frac{d {\bf P}_\alpha}{dt}&=-\nabla_\alpha \left[ \epsilon(\dulR,t) -
\sum_{\alpha=1}^{N_{\rm n}}\frac{1}{2M_\alpha}\frac{\nabla_\alpha^2 \vert\chi(\dulR,t)\vert}{\vert\chi(\dulR,t)\vert}\right]\\
&=-\nabla_\alpha \left[\epsilon(\dulR,t)+\epsilon^{\rm QP}(\dulR,t)\right].
\label{eqn: Newton}
\end{split}
\een
The right-hand side of (\ref{eqn: Newton}) can now be considered as the exact force acting on the 
classical nuclei, since it is derived from the exact nuclear TDSE.
It is a gradient of the sum of the exact TDPES $\epsilon(\dulR,t)$ 
and the additional time-dependent potential
\ben
\epsilon^{\rm QP}(\dulR,t)=-\sum_{\alpha=1}^{N_{\rm n}}\frac{1}{2M_\alpha}
\frac{\nabla_\alpha^2 \vert\chi(\dulR,t)\vert}{\vert\chi(\dulR,t)\vert},
\label{eqn: eQP}
\een
which is referred to as the quantum potential in Bohmian mechanics.

In the previous studies, we propagated multiple classical trajectories according to (\ref{eqn: Newton}) {\it without} 
this Bohmian quantum potential $\epsilon^{\rm QP}(\dulR,t)$, i.e.,
\ben
\begin{split}
\frac{d {\bf P}_\alpha}{dt}=-\nabla_\alpha  \epsilon(\dulR,t) 
\label{eqn: Newton2}
\end{split}
\een
for the field-free nonadiabatic charge-transfer process
\cite{AAYMMG} and 
the laser-induced electron localization
processes in the H$_2^+$ molecule
\cite{localization}.
We found that an ensemble of independent classical nuclear trajectories on $\epsilon(\dulR,t)$
provides dynamics that accurately reproduce the exact nuclear wavepacket dynamics. Here, we 
study whether the same multiple classical trajectory approach can also reproduce strong-field 
processes in which ionization and/or splitting of nuclear density occur/s. In such strong-field 
processes, nuclear quantum effects are significant; in fact, previous studies
\cite{AMG,AMG2,*AMG2C,*AMG2R} have shown that a single classical trajectory cannot yield the molecular 
dissociation via tunneling that occurs under the strong field, even though it is propagated 
by the force of the gradient of the exact TDPES,  $\epsilon(\dulR,t)$.
In the next section, we will propagate multiple classical trajectories by $\epsilon(\dulR,t)$
according to~(\ref{eqn: Newton2}).
We will also calculate the exact quantum potential $\epsilon^{\rm QP}(\dulR,t)$
and propagate multiple classical trajectories by $\epsilon(\dulR,t)+\epsilon^{\rm QP}(\dulR,t)$, i.e., 
(\ref{eqn: Newton}), to study the importance of the force 
from $\epsilon^{\rm QP}(\dulR,t)$.
Note this study demonstrates the role of $\epsilon^{\rm QP}(\dulR,t)$ in
multiple classical-trajectories dynamics 
for the first time.

\section{Results and discussion}

\subsection{Theoretical model}
To study whether the propagation of multiple classical trajectories can reproduce the exact 
quantum nuclear dynamics in strong laser fields, we employ a simplified model of the 
H$_2^+$ molecule, which is the same as that used in previous studies
\cite{SFP4,AMG,AMG2,*AMG2C,*AMG2R,SAMYG,localization,EAM}.
In this model, the dimensionality of the problem is reduced by restricting the motion of the 
nuclei and the electron to the direction of the polarization axis of the laser field
\cite{1DM0, 1DM1, 1dM2, 1DM3}.
In the center-of-mass system, the dynamics of this one-dimensional model of H$_2^+$ 
is governed by the full Hamiltonian 
$\hat{H}(R,r,t)
=\hat{T}_{\rm n}(R)+\hat{T}_{\rm e}(r)+\hat{W}_{\rm nn}(R)+\hat{W}_{\rm en}(R,r)+\hat{v}_{\rm laser}(r,t)
$,
where $R$ is the internuclear distance and $r$ is the electronic coordinate as
measured from the nuclear center of mass.
The kinetic energy terms are $\hat{T}_{\rm n}(R) = -\frac{1}{2\mu_n}\frac{\partial^2}{\partial R^2}$ and, 
$\hat{T}_{\rm e}(r) = -\frac{1}{2\mu_{\rm e}}\frac{\partial^2}{\partial r^2}$, respectively, 
where the reduced mass of the nuclei is given by $\mu_{\rm n}=M_{\rm H}/2$, 
and reduced electronic mass is given by $\mu_{\rm e}=\frac{2M_{\rm H}}{2M_{\rm H}+1}$ ($M_{\rm H}$ 
is the proton mass).
The interactions are soft-Coulomb: $\hat{W}_{\rm nn}(R) = \frac{1}{\sqrt{0.03+R^2}}$,
and $\hat{W}_{\rm en}(R,r) = -\frac{1}{\sqrt{1.0+(r-\frac{R}{2})^2}} -\frac{1}{\sqrt{1.0+(r+\frac{R}{2})^2}}$ 
(and $\hat{W}_{\rm ee} = 0$).
The field is described within the dipole approximation and length gauge, 
as $\hat{v}_{\rm laser}(r,t) = E(t)q_{\rm e}r$,
where the reduced charge $q_{\rm e}=\frac{2M_{\rm H}+2}{2M_{\rm H}+1}$. 
This reduced-dimensional model has proven useful since it allows numerically exact 
solutions to the TDSE while capturing the essential physics in strong-field processes 
such as multiphoton ionization, above-threshold ionization and dissociation, enhanced 
ionization, non-sequential double ionization, and high-harmonic generation
\cite{SFP1,SFP2,SFP3,SFP4,SFP5,SFP6,SFP7,SFP8,SFP9}. 
In this study, we investigate the dynamics of the model H$_2^+$ system under a $\lambda=228$ nm ($5.4$ eV)
UV-laser pulse, which is represented by 
$
E(t)=E_0f(t)\sin(\omega t),
$
with two peak intensities, $I_1 =\vert E_0\vert^2=10^{14}$ W/cm$^2$ and 
$I_2 =\vert E_0\vert^2=2.5\times10^{13}$ W/cm$^2$.
This frequency provides
an energy that is about twice as much as the dissociation 
energy
of the model molecule ($2.88$ eV); thus, dissociation is expected.
The envelope function $f(t)$ is chosen 
such that the field is linearly ramped from zero to its maximum 
strength at $t=7.6$ fs (over 10 optical cycles) and thereafter, held constant
for an additional 15 laser cycles, corresponding to a total simulation time of 
about 19 fs.
The same system and parameters were employed in previous studies
\cite{SFP4,AMG,AMG2,*AMG2C,*AMG2R},
where the important role of the complex coupling between the electronic and nuclear motions in 
these strong-field systems was revealed. In
\cite{AMG,AMG2,*AMG2C,*AMG2R}, in particular, the exact TDPES~(\ref{eqn: tdpes}) 
in these systems was calculated and shown to be a very useful tool for analyzing and 
interpreting the complicated quantum nuclear dynamics in the strong-field processes. 
Here, we will study the possibility of the Bohmian mechanics being established for these 
strong-field processes by using the concept of exact factorization, and whether multiple 
classical trajectories can give the correct quantum nuclear dynamics.

We first calculated the full molecular wavefunction
$\Psi(R,r,t)$
by propagating the full TDSE
\ben
\hat{H}(R,r,t)\Psi(R,r,t)=i\partial_t\Psi(R,r,t)
\label{eqn: fullTDSE}
\een
numerically exactly using the second-order split-operator method
\cite{SPO}.
As the initial state of the time propagation, $\Psi(R,r,t)$ 
was prepared in its ground state by imaginary-time propagation.
\begin{figure}[h]
 \centering
 \includegraphics*[width=1.0\columnwidth]{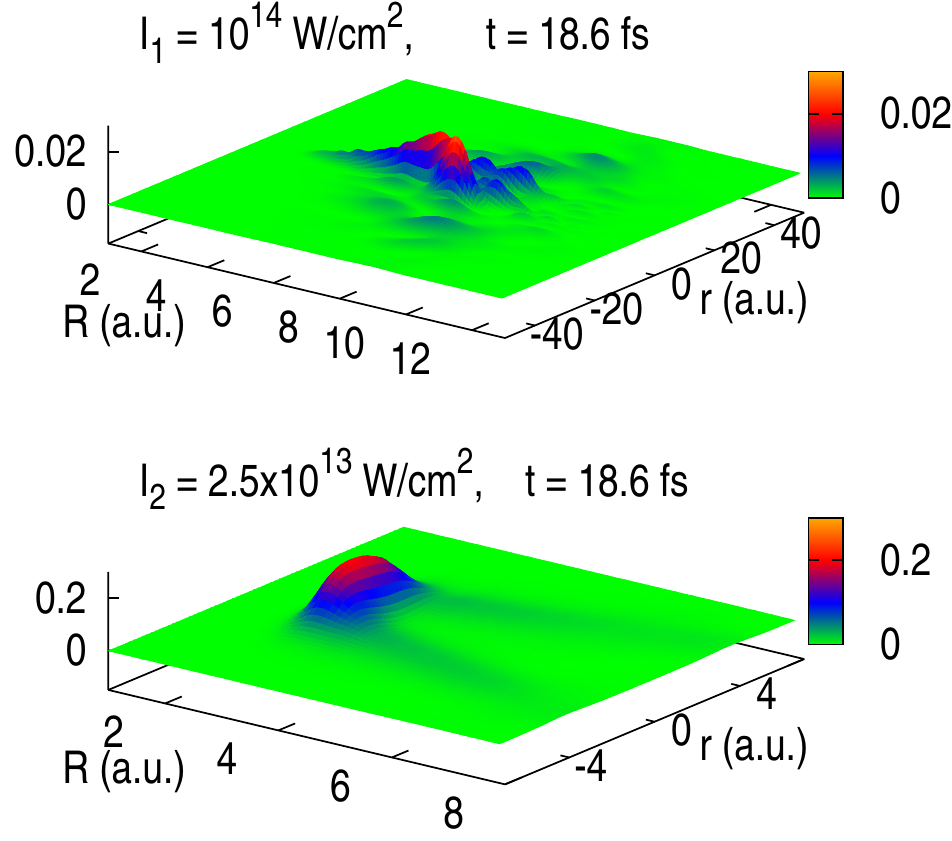}
 \caption{(color online). Electron-nuclear density $|\Psi(R,r,t)|^2$
 for the model H$_2^+$ molecule after the 24th optical cycle 
 ($t=18.6$ fs) in a $\lambda=228$ nm laser field. (Upper panel:
 higher-intensity case ($I_1 =10^{14}$ W/cm$^2$).
 Lower panel: lower-intensity case ($I_2 =2.5\times10^{13}$ W/cm$^2$.) 
 In atomic units.}
 \label{fig:Fig1}
\end{figure}
In Fig.~\ref{fig:Fig1}, the electron-nuclear density $|\Psi(R,r,t)|^2$
at $t=18.6$ fs (after the 24th optical cycle) 
is shown for both the higher-intensity case ($I_1 =10^{14}$ W/cm$^2$) (upper panel)
and the lower-intensity case ($I_2 =2.5\times10^{13}$ W/cm$^2$) (lower panel).
These indicate the probability of finding an electron at position $r$
and the nuclear separation at position $R$ at $t=18.6$ fs for each case.
In the upper panel, it is observed that 
$|\Psi(R,r,t)|^2$ at $t=18.6$ fs exists at larger $R$ compared to the 
expectation value at the ground state $\langle R \rangle(t=0)=2.65$ a.u.,
indicating that dissociation occurred. 
We also observe large streaks of
$|\Psi(R,r,t)|^2$ in both
negative and positive $r$ 
directions, which shows that a considerable ionization occurred in this higher-intensity case. 
Therefore, dissociation occurs here via the Coulomb-explosion mechanism, as already discussed in previous studies
\cite{SFP4,AMG,AMG2,*AMG2C,*AMG2R}.
However, in the lower-intensity case (lower panel in Fig.~\ref{fig:Fig1}),
different dynamics occurred: a small amount of $|\Psi(R,r,t)|^2$
exists in the region larger than $\langle R \rangle(t=0)=2.65$ a.u.,
but a large part of it remains in the ground-state position around $\langle R \rangle(t=0)$.
Therefore, a splitting of probability density occurred here and a small amount of  $|\Psi(R,r,t)|^2$ 
went to dissociation. It is also seen that the probability of ionization is very low, and hence, 
dissociation occurred predominantly via the photodissociation channel
(H$^+_2\rightarrow$H$+$H$^+$)
\cite{SFP4,AMG,AMG2,*AMG2C,*AMG2R}. 
Previous studies
\cite{SFP4,AMG,AMG2,*AMG2C,*AMG2R} have shown that this lower-intensity case ($I_2 =2.5\times10^{13}$ W/cm$^2$) 
represents a particularly challenging system when we consider simulating it by using the 
approximated method. The Ehrenfest and time-dependent Hartree methods could not reproduce the 
probability of dissociation of this system. A more sophisticated correlated time-dependent variational approach
\cite{SFP4}
succeeded in giving the dissociation probability to some degree, but still could not reproduce 
the nuclear density dynamics well. The exact TDPES in this system
\cite{AMG,AMG2,*AMG2C,*AMG2R} 
provided a clear picture that explains the difficulty in the simulation of these nuclear dynamics: 
it was shown that the quantum tunneling through the TDPES occurs, which is difficult to reproduce 
by the approximated methods. The question arises as to whether we get the force acting on classical 
nuclei that gives the correct dynamics in this system if we formulate Bohmian mechanics 
in the exact-factorization framework.

\subsection{Multiple classical trajectory dynamics on the exact TDPES + the exact quantum potential}

\begin{figure}[]
 \centering
 \includegraphics*[width=1.0\columnwidth]{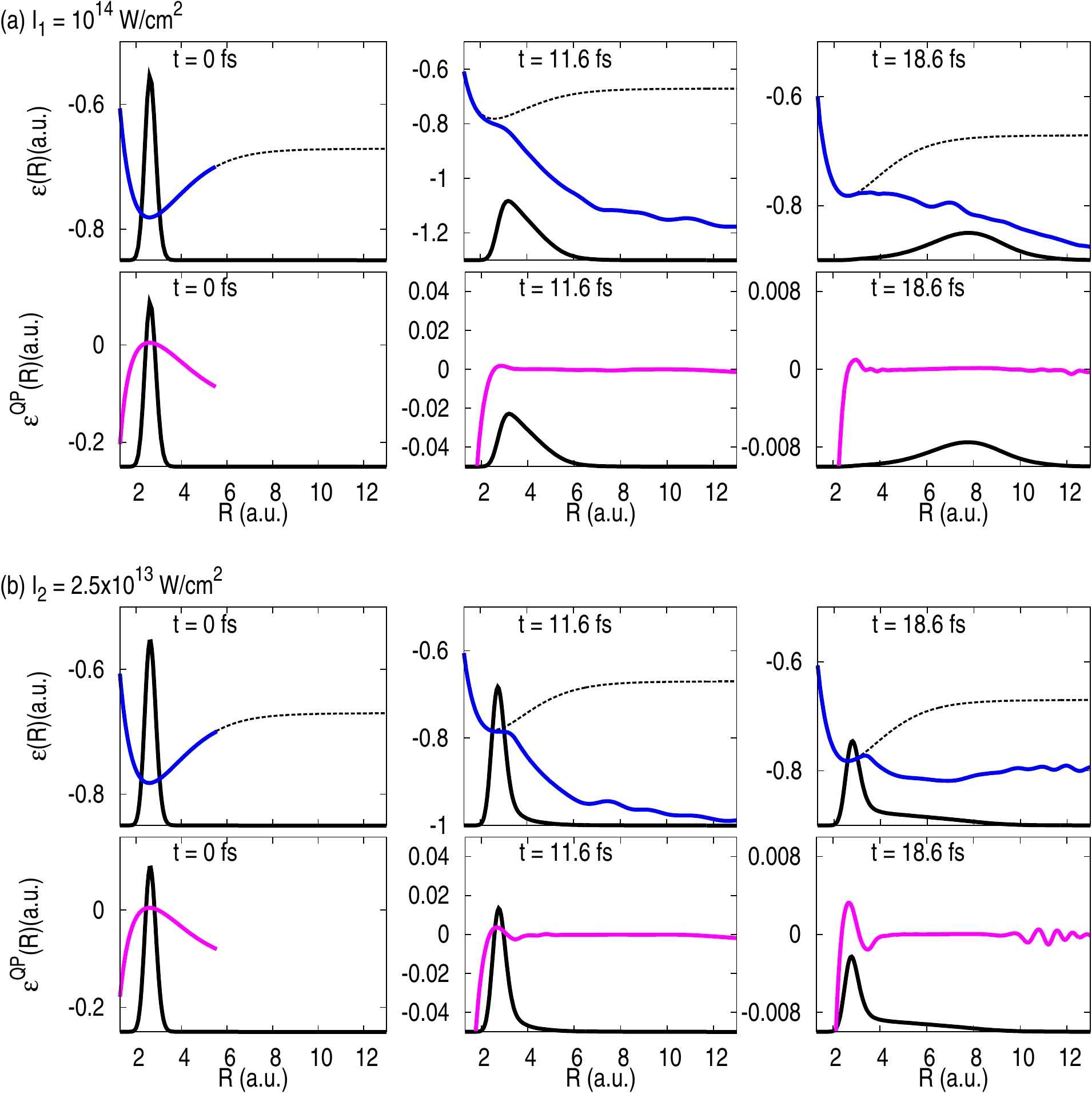}
 \caption{(color online). (a) Top panel: Snapshots of the exact TDPES $\epsilon(R,t)$ (blue or dark gray solid lines)
 and nuclear density (black solid lines) at times indicated,
 for H$^+_2$ subjected to the laser-field ($\lambda=228$ nm) with the peak intensity $I_1 =10^{14}$ W/cm$^2$.
For reference, the ground-state BO potential-energy surface (black dashed lines) is also shown.  
 Bottom panel: Snapshots of the exact quantum potential $\epsilon^{\rm QP}$ (pink or light gray solid lines) 
 for the same system as in the top panels
 at the same times. Nuclear density is again shown (black solid lines).
 (b) Same as (a) but for the case of lower intensity ($I_2 =2.5\times10^{13}$ W/cm$^2$).}
 \label{fig:Fig2}
\end{figure}

To answer the questions raised in the previous sections, we now present our results. We begin 
with the calculation of the exact TDPES
 $\epsilon(R,t)$~(\ref{eqn: tdpes}) and 
the quantum potential $\epsilon^{\rm QP}(R,t)$~(\ref{eqn: eQP}) that appear in the exact quantum Hamilton-Jacobi
equation~(\ref{eqn: QHJeqn}). 
Since we already have numerically exact $\Psi(R,r,t)$ at each time step obtained 
by propagating the full TDSE~(\ref{eqn: fullTDSE}), 
we can easily calculate the TDPES $\epsilon(R,t)$ by fixing the gauge 
\cite{AMG2,*AMG2C,*AMG2R}
and $\epsilon^{\rm QP}(R,t)$ with $|\chi(R,t)|=\sqrt{\int{|\Psi(R,r,t)|^2 dr}}$. 
In Fig.~\ref{fig:Fig2} (a) and (b), we show 
snapshots of the exact TDPES $\epsilon(R,t)$ (blue or dark gray solid lines)
and the quantum potential $\epsilon^{\rm QP}(R,t)$ (pink or light gray solid lines)
at times indicated for the system with peak intensities $I_1 =10^{14}$ W/cm$^2$ (Fig.~\ref{fig:Fig2} (a))
and $I_2 =2.5\times10^{13}$ W/cm$^2$ (Fig.~\ref{fig:Fig2} (b)), respectively.
The nuclear density at each time is also shown (black solid lines). Note that the TDPES have 
already been reported in previous studies
\cite{AMG,AMG2,*AMG2C,*AMG2R},
and here we show the quantum potential for the first time. The TDPES in each system shows the 
characteristic feature of the time-dependent potentials that the nuclear wavepacket experiences. 
In the Coulomb explosion case (Fig.~\ref{fig:Fig2}(a)), 
the wells in the TDPES $\epsilon(R,t)$ 
that confine the wavefunction in the ground state flatten out as the laser is switched on, causing 
the nuclear density to spill out to larger separations. However, in the photodissociation (without ionization) case
 (Fig.~\ref{fig:Fig2} (b)),
the well in the $\epsilon(R,t)$ 
exists at all times, and thus, the nuclear density can leak out from it only by tunneling.

\begin{figure*}[]
 \centering
 \includegraphics*[width=2.0\columnwidth]{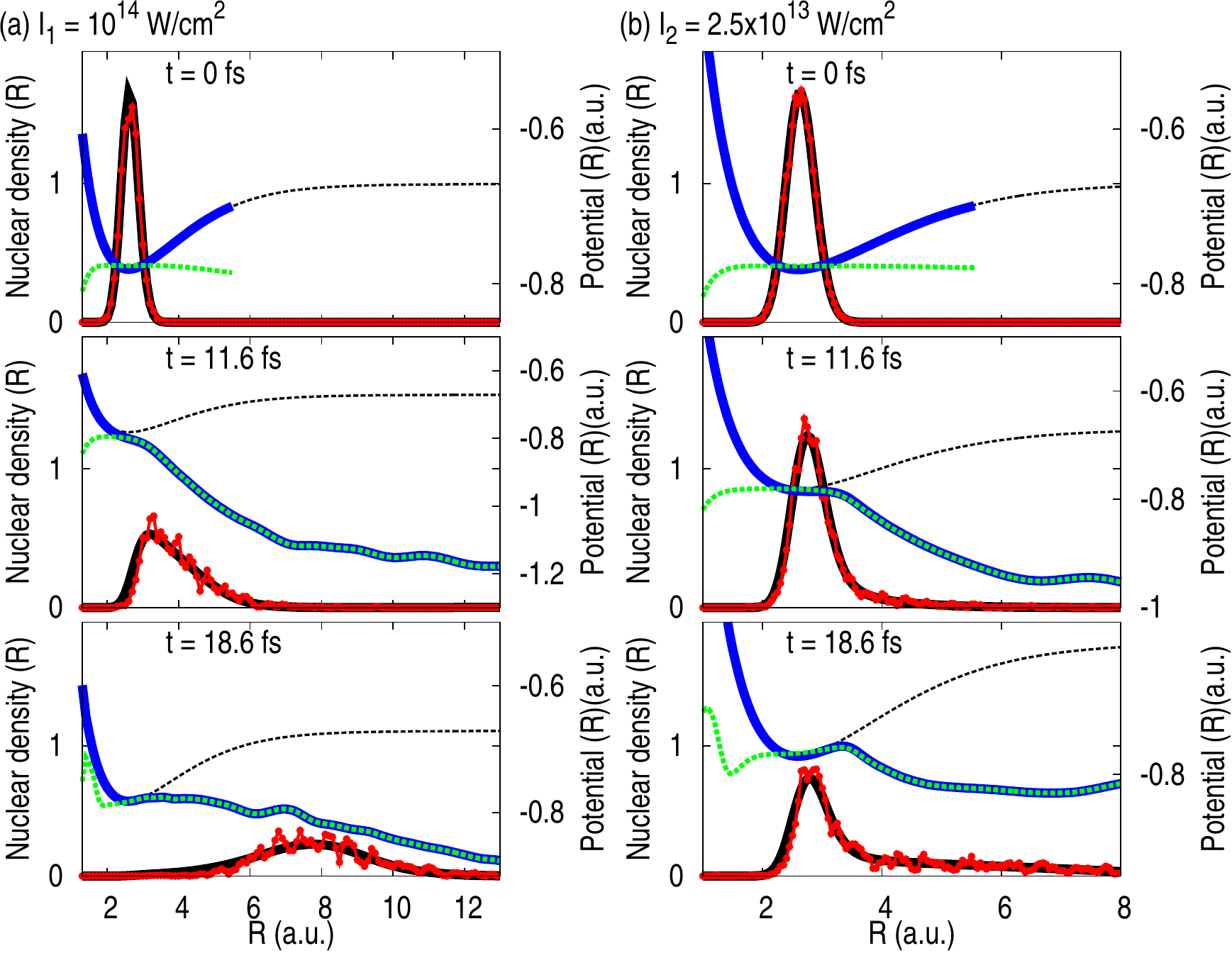}
 \caption{(color online). (a) Snapshots of the nuclear density reconstructed as a histogram 
 from the distribution of the classical positions obtained by solving Newton's 
 equation~(\ref{eqn: Newton}) 
 for H$^+_2$ subjected to the laser field  ($\lambda=228$ nm) with peak intensity 
 $I_1 =10^{14}$ W/cm$^2$ at indicated times
 (red or gray linepoints). Exact nuclear density is also shown (black solid line). Green or light gray dashed line 
 indicates the sum of the exact 
 TDPES and quantum potential ($\epsilon(R,t) + \epsilon^{\rm QP}(R,t)$) 
 whose gradient is the force used to obtain the red or gray linepoints. Blue or dark gray solid line is
  $\epsilon(R,t)$ and black dashed line is the ground-state BO potential-energy surface. 
  (b) Same as (a), but for the case of lower intensity
   ($I_2 =2.5\times10^{13}$ W/cm$^2$).
 }
 \label{fig:Fig3}
\end{figure*}

By comparing the upper and lower panels in both laser cases, we find that
$\epsilon^{\rm QP}(R,t)$ 
has non-negligible structures, especially at earlier simulation times. It is found that
$\epsilon^{\rm QP}(R,t)$ actually has an effect that flattens the well in $\epsilon(R,t)$, 
thus helping the nuclei delocalize against the force from the confining potential 
at the initial time and leak out from it.
Especially at 
the initial time when the wavefunction is in its ground state, the shape of
$\epsilon^{\rm QP}(R,t=0)$ is exactly opposite to that of TDPES $\epsilon(R,t=0)$
as seen in the left-hand panels of Figs.~\ref{fig:Fig2} (a) and (b).
This is understood by the fact that the ground state of a one-dimensional wavefunction 
is expressed by a real function multiplied by a complex constant, i.e., 
 $\chi(R)=|\chi(R)|e^{ia}$ where $a$ is a real constant.
Since the exact nuclear Schr\"{o}dinger equation in the ground state is written as
\cite{Gphase1,staticEF}
\ben
 \left[- \frac{1}{2\mu_n}\frac{d^2}{dR^2} +
 \epsilon(R)\right]\chi(R)
 =E \chi(R) \label{eqn: static nuclear eqn}
 \een
(where $E$ is the total energy of the system in the ground-state), the quantum potential 
at the initial time can be written as
\ben
\epsilon^{\rm QP}(R)=\frac{-\frac{1}{2\mu_{\rm n}}\frac{d^2}{dR^2}|\chi(R)|}{|\chi(R)|}=E-\epsilon(R).
\label{eqn: static eQP}
\een
Thus, $\epsilon^{\rm QP}(R)$ shows the opposite curvature to the exact TDPES $\epsilon(R)$. 
We now demonstrate that the effect of $\epsilon^{\rm QP}$ 
on the dynamics is significant, especially for the low-intensity case, as it causes the 
tunneling of the classical nuclei.

Having the exact TDPES $\epsilon(R,t)$ and the quantum potential $\epsilon^{\rm QP}(r,t)$ at each time step,
we now solve the {\it exact} quantum Hamilton-Jacobi equation~(\ref{eqn: QHJeqn}) 
for these two systems by propagating multiple classical trajectories according to Newton's equation
~(\ref{eqn: Newton}).
We first show the results for the higher-intensity
 ($I_1 =10^{14}$ W/cm$^2$) case.
We propagate 1000 trajectories according to Eq.~(\ref{eqn: Newton}), where the initial 
positions $R$ are sampled from the 
initial (ground-state) nuclear density $N(R,t=0)=\int{|\Psi(R,r,t=0)|^2 dr}$, 
which was obtained in the previous TDSE calculation, and the initial momentum is set to zero since
$\frac{d}{dR}S(R,t)$
is zero at the initial time.
In Fig.~\ref{fig:Fig3} (a), 
the nuclear density reconstructed as a histogram from the distributions of classical positions evolving on
$\epsilon(R,t)+\epsilon^{\rm QP}(R,t)$ 
for the system with the peak intensity $I_1 =10^{14}$ W/cm$^2$ is shown
as red (or gray) linepoints.
The exact nuclear density obtained from the full TDSE~(\ref{eqn: fullTDSE})
is also shown as a black solid line. By comparing the red (or gray) linepoints and black line, we find that the 
nuclear density obtained from the multiple classical trajectories yields the exact quantum nuclear 
dynamics in this strong-field process: the characteristic Coulomb-explosion dynamics of the 
quantum nuclei are reproduced by the ensemble of classical trajectories. This result indicates 
that even strong-field processes in which ionization occurs can be simulated, in principle, by 
the MQC approximation method. We also show
$\epsilon(R,t)+\epsilon^{\rm QP}(R,t)$ (green or light gray dashed line) and $\epsilon(R,t)$ (blue or dark gray solid line). 
By comparing these, the discussion above is confirmed: 
$\epsilon^{\rm QP}$ plays a role to
flatten the well in $\epsilon(R,t)$, which initially confines the wavepacket 
to the equilibrium position, thus enhancing dissociation. 
Here, in the higher-intensity
($I_1 =10^{14}$ W/cm$^2$) 
case, together with the TDPES that gives a strong repulsive force that reflects ionization, almost 
the entire nuclear density leaks out from the well and moves to the dissociation. We observe that, 
when the nuclear density moves outside the well, the effect of
$\epsilon^{\rm QP}$ 
becomes very small. 
Thus, in this higher-intensity case, the quantum potential affects the dynamics 
only during the earlier time of the propagation when the nuclear density is about to leak out from the well.

Next, we turn to the lower-intensity case ($I_2 =2.5\times10^{13}$ W/cm$^2$),
where tunneling of the nuclear wavepacket occurs and causes the difficulty to be simulated by the approximated method
\cite{SFP4,AMG,AMG2,*AMG2C,*AMG2R}.
We propagate 2,000 trajectories according to
~(\ref{eqn: Newton}) with
$\epsilon(R,t)$ plus $\epsilon^{\rm QP}(R,t)$ 
calculated above. Snapshots of the nuclear density reconstructed as a histogram from the distributions 
of classical positions are plotted as red (or gray) linepoints in Fig.~\ref{fig:Fig3} (b). 
Comparison with the exact nuclear density (black solid line in the same figure) again shows excellent agreement. 
Multiple classical trajectories evolving on
 $\epsilon(R,t)+\epsilon^{\rm QP}(R,t)$ 
perfectly reproduce the splitting of the exact nuclear density, i.e., long-reaching tails of the exact 
nuclear density are correctly reproduced. Thus, it is found that the strong-field photodissociation 
dynamics can also be simulated, in principle, by multiple classical trajectories when their motion 
is driven by the correct force, i.e., the gradient of
$(\epsilon(R,t)+\epsilon^{\rm QP}(R,t))$. 
In the same figure,
$\epsilon(R,t)+\epsilon^{\rm QP}(R,t)$  and $\epsilon(R,t)$ 
are also plotted.
Similar to the higher-intensity case,
$\epsilon^{\rm QP}(R,t)$ flattens the well in $\epsilon(R,t)$; thus,
it allows the nuclear density to escape from the confining potential.   
This explains how quantum tunneling is reproduced by classical-trajectory dynamics: the quantum potential 
is responsible for it. Here, a large part of the nuclear density remains in the ground-state position 
and the exact TDPES
$\epsilon(R,t)$ 
has a confining well at all times. However, the quantum potential
$\epsilon^{\rm QP}(R,t)$ 
has the opposite curve to
$\epsilon(R,t)$ 
around the equilibrium position. Thus, in total,
$\epsilon(R,t)+\epsilon^{\rm QP}(R,t)$  
always shows a somewhat flattened structure and some classical nuclei can experience the force to escape 
the well and go to the dissociation. Note that once the nuclei leak outside the well, the quantum potential 
$\epsilon^{\rm QP}(R,t)$
has an almost negligible effect on their dynamics.

\begin{figure}[]
 \centering
 \includegraphics*[width=1.0\columnwidth]{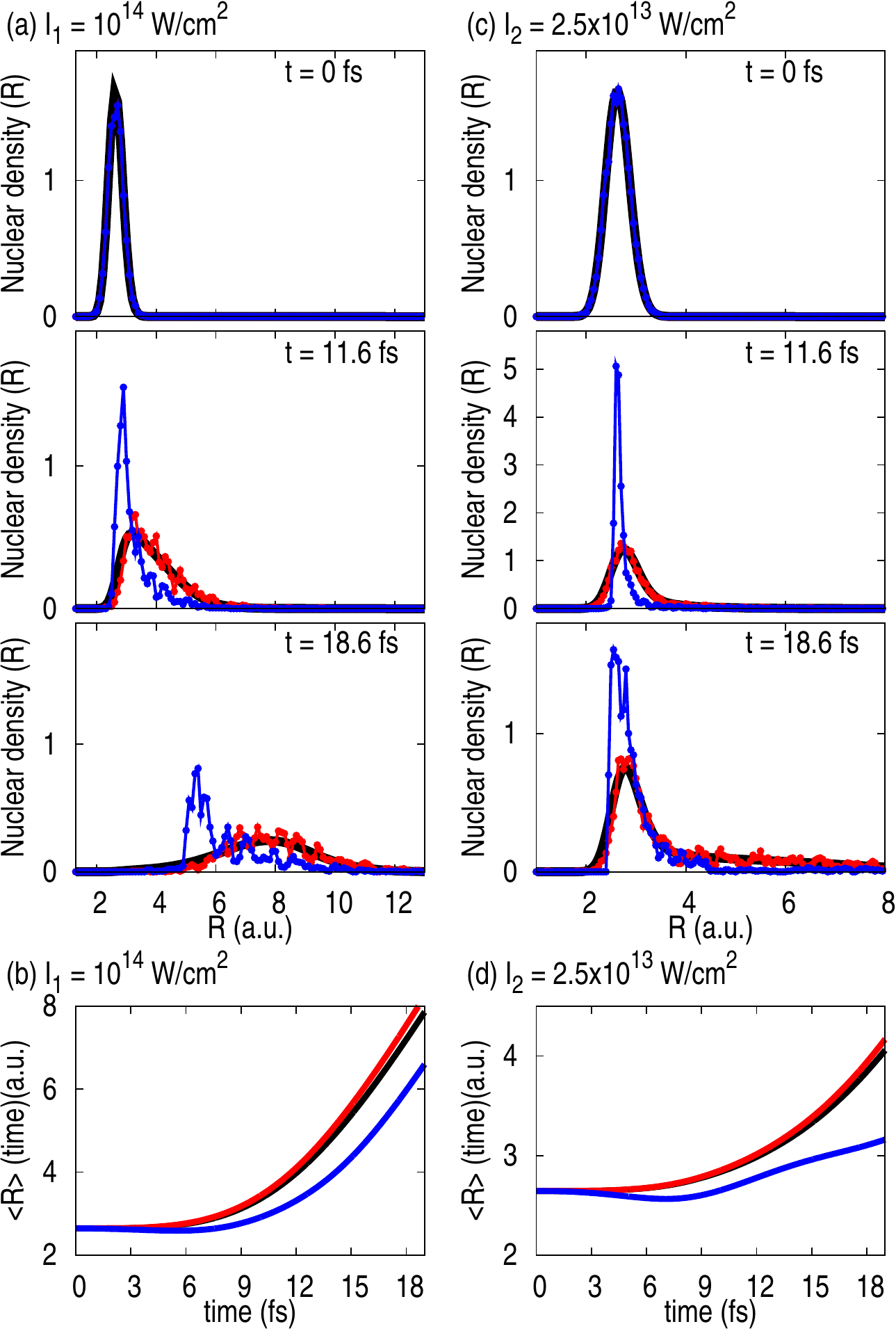}
 \caption{(color online). (a) Snapshots of the nuclear density reconstructed as a histogram from the distribution of 
 the classical positions at indicated times for the case where
  H$^+_2$ is subjected to the higher-intensity ($I_1 =10^{14}$ W/cm$^2$) laser field ($\lambda=228$ nm).
 The red or light gray linepoints represent the density obtained by propagating classical trajectories on 
 $\epsilon(R,t)+\epsilon^{\rm QP}(R,t)$
 , while the blue or dark gray linepoints indicate the results obtained by propagating classical trajectories 
 only on $\epsilon(R,t)$.
 Exact nuclear density is also shown as a black solid line. (b) Time evolution of the mean inter-nuclear distance
  $\langle R\rangle(t)$ 
  obtained for the same system as in (a) from three different simulations: propagation of classical trajectories on
   $\epsilon(R,t)+\epsilon^{\rm QP}(R,t)$ (red or light gray solid line), 
   propagation of classical trajectories only on
  $\epsilon(R,t)$ (blue or dark gray solid line), and the exact TDSE solution (black solid line). 
  (c/d) Same as (a/b) but for the case of lower intensity ($I_2 =2.5\times10^{13}$ W/cm$^2$).  }
 \label{fig:Fig4}
\end{figure}

Finally, to confirm if the above discussion about the role of the quantum potential
$\epsilon^{\rm QP}(R,t)$  
is correct, we compute the multiple classical trajectory dynamics without taking account of the quantum 
potential, i.e., propagated according to
~(\ref{eqn: Newton2}). 
In Fig.~\ref{fig:Fig4} (a) and (c), 
we show snapshots of the calculated dynamics: the blue (or dark gray) linepoints indicate the nuclear density 
reconstructed as a histogram from the distribution of the classical positions evolving only on
$\epsilon(R,t)$.
((a) shows the results for the higher-intensity case and (c) is for the lower-intensity one). 
As a reference, the densities obtained from the multiple classical trajectories on
$\epsilon(R,t)+\epsilon^{\rm QP}(R,t)$ (red or light gray linepoints)
and the exact nuclear density (black solid line) are also shown. To see the difference between these 
results more clearly, in Figs.~\ref{fig:Fig4} (b) and (d),
we also plot the time evolution of the mean inter-nuclear distance
$\langle R\rangle(t)$ 
obtained from each calculation ((b) is for the higher-intensity case and (d) is for the 
lower-intensity one). Exact results (black solid line) are obtained by
$\langle R\rangle(t)=\langle\Psi(R,r,t)|\hat{R}|\Psi(R,r,t)\rangle$  
while $\langle R\rangle(t)$ 
from classical trajectories (red or light gray solid line indicates the dynamics on
$\epsilon(R,t)+\epsilon^{\rm QP}(R,t)$
and blue or dark gray solid line indicates the dynamics only on
$\epsilon(R,t)$) are obtained as
$\langle R\rangle(t)=\frac{1}{N_{\rm traj}}\sum^{N_{\rm traj}}_{I=1}R_I(t)$, where
$N_{\rm traj}$ is the total number of trajectories and $R_I(t)$ is the distance at time $t$ of each trajectory.
Comparison between the blue (or dark gray) and red (or light gray) lines clearly reveals the failure in the dynamics propagated only by the force from
$\epsilon(R,t)$. 
In the higher-intensity case, the dynamics driven only by 
$\epsilon(R,t)$ 
succeeded in reproducing the dissociation ones, but their speed is lower than the exact result; 
the shape of the nuclear density (blue or dark gray) is more localized in the smaller
$R$ 
region than the exact density. This is easily understood because the quantum potential has the 
effect of flattening the well of the ground-state potential, as shown above, and its absence 
makes the nuclear density likely to be trapped at the equilibrium position, leading to slower dissociation.

In the lower-intensity case, failure in the dynamics propagated only by the force from
$\epsilon(R,t)$ 
is more noticeable: the majority of the nuclear density shown as the blue (or dark gray) curve is trapped by the 
confining potential well at the equilibrium position, and only a very small fraction leaks out 
to the dissociation, even in the final snapshot
($t=18.6$ fs). The time evolution of  $\langle R\rangle(t)$ in Fig.~\ref{fig:Fig4} (d)
clearly reflects this: the speed of the dissociation in the dynamics
propagated only by $\epsilon(R,t)$ 
is much lower than that in the exact dynamics. This result is also supported by the above 
analysis, which shows that the quantum potential would have flattened the well and caused 
tunneling if it had been included in the force. From these results, we can conclude that 
the quantum potential plays a non-negligible role in reproducing the exact quantum nuclear 
dynamics in the strong-field processes studied here by propagating an ensemble of classical 
trajectories according to the Newton's equation~(\ref{eqn: Newton}) 
derived from Bohmian mechanics in the exact nuclear TDSE~(\ref{eqn: exact nuclear eqn}). 
This suggests that careful assessment of its effect is required when
developing the MQC method based on the exact nuclear TDSE~(\ref{eqn: exact nuclear eqn}), especially when we aim to develop the method 
for strong-field processes.

\section{Conclusions}

In this paper, we have demonstrated that the propagation of multiple classical trajectories can 
reproduce quantum nuclear dynamics even in strong-field processes when they are propagated by 
Newton's equation with the force determined from the gradient of the exact TDPES {\it plus} the exact 
quantum potential, which is defined in Bohmian mechanics in the exact-factorization framework. 
We employed a one-dimensional
H$_2^+$ 
model system subject to two different-intensity laser fields, which give rise to different 
types of dissociation dynamics, and showed that both processes can be simulated by multiple 
classical trajectory dynamics. We found that the exact quantum potential has a non-negligible 
effect on the classical dynamics: its force accelerates the classical nuclei to overcome the 
confining well of the ground-state potential-energy surface, and causes tunneling of the 
nuclear density in the lower-intensity case.

The results here offer important knowledge for developing the MQC algorithm for coupled 
electron-nuclear dynamics based on the exact-factorization approach, especially when one 
wants to develop it for strong-field processes. Although it is a challenging task to develop 
suitable approximations to the quantum Hamilton-Jacobi equation
(\ref{eqn: QHJeqn}) 
and make it solvable on-the-fly, our results are encouraging since they show that quantum 
nuclear dynamics can be simulated, in principle, by classical trajectories if the force 
is properly prepared, and the existence of such exact forces is proved. In this study, we 
explored the case where the gauge is fixed such that the vector potential in the exact nuclear TDSE 
vanishes. It is also desirable to formulate Bohmian mechanics in a more general gauge, i.e., 
where the vector potential exists and the gauge-dependent part of the TDPES vanishes. This would 
provide a more general basis for developing an MQC method that can yield the correct quantum 
nuclear-density dynamics in various non-adiabatic situations including strong-field processes.

{\it Acknowledgments:}
This study was supported by JSPS KAKENHI Grant No. 16K17768.

\bibliography{./exactforce}

\end{document}